\documentclass[
superscriptaddress,
     amsmath,amssymb,
     aps,
    pra,
    onecolumn,
    ]{revtex4-2}
\usepackage{tcolorbox}
\usepackage{adjustbox}
\usepackage{xcolor}
\usepackage{graphicx} 
\usepackage{dcolumn} 
\usepackage{bm} 
\usepackage{braket} 
\usepackage{amsthm}
\usepackage{bbm}
\usepackage[normalem]{ulem}
\usepackage[ruled]{algorithm2e}
\usepackage{booktabs}
\usepackage{mathtools}
\usepackage{tabularray}

\newtheorem*{theorem-non}{Theorem}
\newtheorem*{lemma-non}{Lemma}
\newtheorem*{corollary-non}{Corollary}
\newtheorem*{proposition-non}{Proposition}

\newcommand{\omni}{\textsf{OmniQEC}\xspace}

 \usepackage{appendix} 
 \usepackage[colorlinks=true,
            citecolor=NatureTeal,
            linkcolor=NatureRed,
            urlcolor=NatureRed]{hyperref}

\usepackage{fontawesome5}
 
\definecolor{NatureTeal}{HTML}{0F6B68}
\definecolor{NatureBlue}{HTML}{1F4E79}
\definecolor{NatureRed}{HTML}{A23B3B}

\begin{document}

\title{\omni: discovering practical quantum error-correcting codes by an AI scientist}

\author{Ge Yan}
\affiliation{College of Computing and Data Science, Nanyang Technological University, Singapore}
\author{Shanchuan Li}
\affiliation{College of Computing and Data Science, Nanyang Technological University, Singapore}
\affiliation{Department of
Electrical Engineering and Computer Science, Tokyo University
of Agriculture and Technology, Koganei, Tokyo, Japan}
\author{Pengyue Ma}
\affiliation{College of Computing and Data Science, Nanyang Technological University, Singapore}
\author{Qixin Zhang}
\affiliation{College of Computing and Data Science, Nanyang Technological University, Singapore}
\author{Pingchuan Ma}
\affiliation{Zhejiang University of Technology, China}
\author{Jianping Wang}
\affiliation{Department of Computer Science, City University of Hong Kong, Hong
Kong, China}
\author{Min-Hsiu Hsieh}
\affiliation{Hon Hai (Foxconn) Research Institute, Taipei, Taiwan}
\author{Yuxuan Du}
\email{yuxuan.du@ntu.edu.sg}
\affiliation{College of Computing and Data Science, Nanyang Technological University, Singapore}
\affiliation{School of Physical and Mathematical Science, Nanyang Technological University, Singapore}
    
\begin{abstract}
Quantum error correction (QEC) is indispensable for scalable fault-tolerant quantum computing. However, discovering QEC codes that remain effective is challenging, as logical performance depends on the interplay between code structure, hardware, syndrome extraction, and decoding, which often impose competing requirements. Here we introduce \omni, an efficient AI scientist for discovering QEC codes suited to deployment on modern quantum processors. \omni formulates QEC design as an iterative discovery process in which an orchestrator, implemented by advanced large language models (LLMs), coordinates code generation, code-level screening, syndrome-extraction synthesis, and decoder-based circuit evaluation. At its core, \omni combines a self-evolving reasoning mechanism with a slow--fast synergistic workflow: a fast loop explores candidates using inexpensive code-level proxies, whereas a slow loop performs physically grounded circuit-level evaluation and feeds the resulting evidence back into the search. We evaluate \omni across four qLDPC construction families, three LLM backends, and $14$ total-physical-qubit budgets per backend. The discovered codes show steadily improving logical-error suppression with increasing physical-qubit budgets and outperform the BB codes with $[\![72,12,6]\!]$ and $[\![144,12,12]\!]$ under complete-implementation budgets of 98 and 240 physical qubits, respectively. The discovered codes are hardware-friendly and may be of independent interest for practical QEC implementation. These findings pave the way towards LLM-assisted QEC discovery grounded in physically informed code--circuit--decoder co-design.
\end{abstract}

\maketitle

\begin{figure*}
    \centering
    \includegraphics[width=\linewidth]{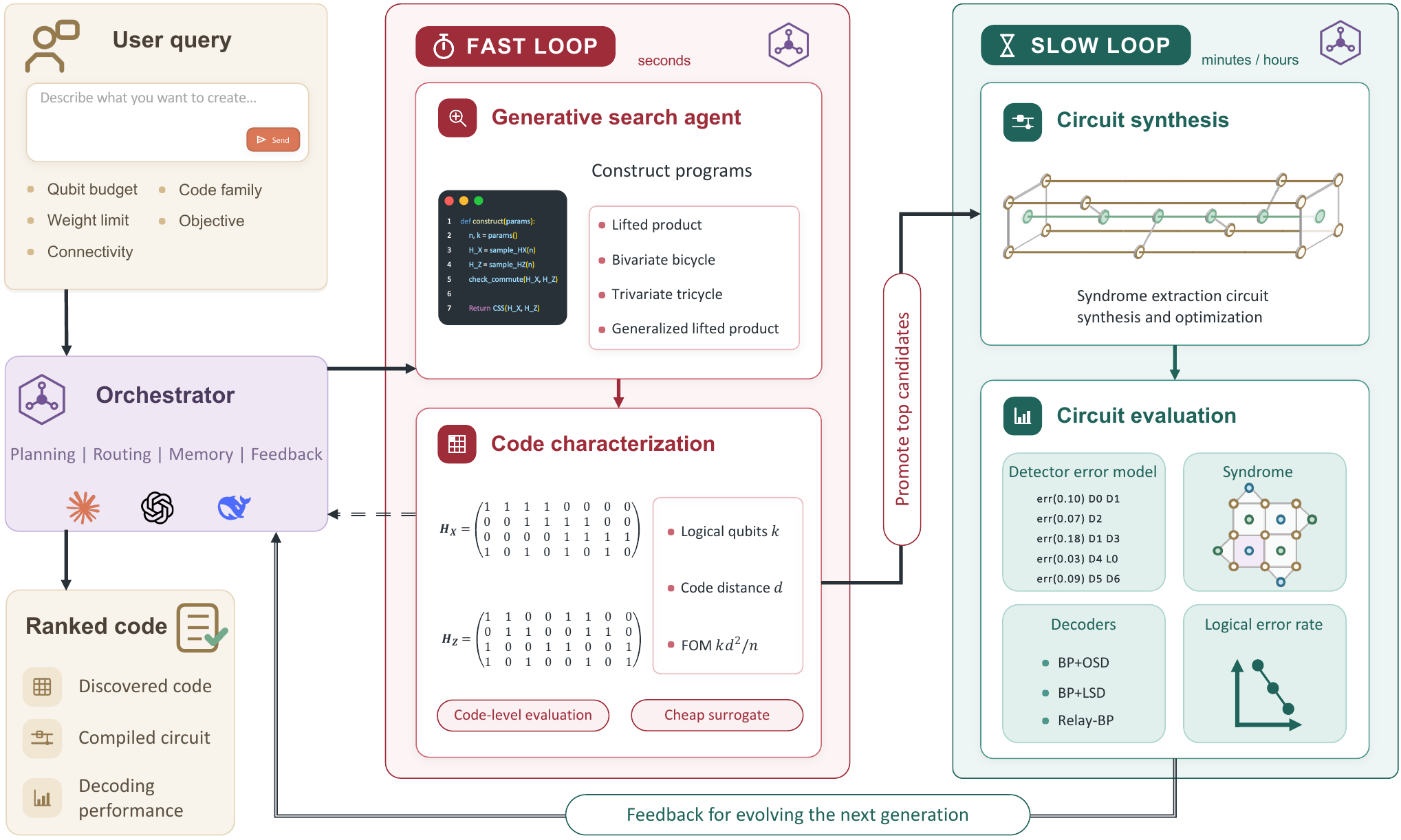}
   \caption{\textbf{Schematic illustration of \omni.} A user specifies a QEC design objective, such as the target code family, optimization criterion, physical-qubit budget, connectivity, native gate set, and available ancilla resources. The LLM-based orchestrator translates this objective into a structured discovery task, maintains the self-evolving discovery memory, and routes information across a dual-loop workflow. In the fast loop, the code-search agent proposes executable code-construction programs, which are converted into candidate QEC codes and screened by the code-characterization module using code-level properties, including validity, logical-qubit number, code distance, and the code-level proxy such as figure of merit (FOM). Promising candidates are promoted to the slow loop, where the circuit-synthesis module compiles validated codes into syndrome-extraction circuits and the circuit-evaluation module estimates circuit-level logical error rates using noisy sampling and selected decoders. The resulting circuit-level evidence is returned to the orchestrator as slow-loop feedback and incorporated into the discovery memory, enabling subsequent iterations to generate improved candidates. The workflow outputs a ranked package containing reproducible code-construction programs, discovered codes, compiled circuits, and decoding performance.}
    \label{fig:scheme}
\end{figure*}

\section{INTRODUCTION} 
Quantum error correction (QEC) is the foundation for fault-tolerant quantum computation (FTQC)~\cite{gottesman2009introduction,terhal2015quantum,roffe2019quantum}. Its central role has driven a broad landscape of code families. Representative codes include topological surface and toric codes for geometrically local protection~\cite{kitaev2003fault,dennis2002topological,fowler2012surface}, color codes with transversal operations~\cite{bombin2006topological,lacroix2025scaling}, subsystem and Floquet codes for simplified syndrome extraction~\cite{hastings2021dynamically,davydova2023floquet,gidney2021fault,cao2026maximum}, high-rate qLDPC codes for low-overhead memory with sparse checks and flexible connectivity~\cite{tillich2014quantum,panteleev2021quantum,breuckmann2021balanced,panteleev2022asymptotically,bravyi2024high}, and concatenated or bosonic codes for hierarchical or mode-level protection~\cite{goto2024high,ruiz2025ldpccat}. Despite this diversity, only a small subset of these constructions has reached modern quantum processors, including surface-code experiments~\cite{google2023suppressing,google2025quantum} and emerging colour-code logic~\cite{lacroix2025scaling} on superconducting devices, encoded logical circuits on reconfigurable atom arrays~\cite{bluvstein2024logical,xu2024constant}, and real-time fault-tolerant logical operations with trapped ions~\cite{egan2021fault,ryananderson2021realization}. This gap reflects a conventional code-centric paradigm that separates algebraic construction from the constraints of practical deployment~\cite{bravyi2010tradeoffs}. As hardware platforms progress towards early FTQC~\cite{katabarwa2024early}, flexible architectures expand the experimentally accessible qLDPC design space~\cite{zhou2025opportunities}. This expanded design space, however, places practical and high-performing QEC codes increasingly beyond the reach of purely expert-driven search~\cite{dennis2002topological,panteleev2021quantum,panteleev2022asymptotically,bravyi2024high}.

Recent advances in artificial intelligence (AI) have initiated early attempts to automate the discovery of practical QEC codes. Existing approaches broadly fall into three categories. One class uses reinforcement learning to search over code constructions with reward functions defined by favorable properties~\cite{nautrup2019optimizing,olle2024simultaneous,su2025discovery,freire2025optimizing,he2025discovering,du2025artificial}. A second uses Bayesian optimization to guide candidate selection~\cite{chengyu2026bayesian}, whereas a third builds on large language model (LLM)-guided program search to generate and iteratively refine executable code constructions~\cite{cruz2026evolutionary,he2026co,liu2026large}. Nevertheless, all paradigms remain predominantly \textit{code-centric}, ranking candidates using algebraic quantities such as $[\![n,k,d]\!]$ or inexpensive proxies derived from them~\cite{calderbank1996good,steane1996multiple,gottesman1997stabilizer}. In other words, they fail to capture the practical determinants of QEC performance, such as hardware constraints, syndrome extraction, and decoding. As a result, the designed QEC code that appears promising at the code level may deliver poor logical performance in practice~\cite{higgott2023improved,zhang2026optimal,viszlai2026prophunt}. This raises a central question: \begin{center}
    \textit{how can QEC discovery move beyond code-level optimization towards practical performance?}
\end{center} 

Here we introduce \omni, an AI scientist for discovering practical QEC codes by directly optimizing the circuit-level logical error rate (LER), a standard metric of implementation-level QEC performance~\cite{google2023suppressing,google2025quantum}. Conceptually, \omni casts QEC code discovery as an autonomous generate–evaluate–refine process, in which an LLM-based orchestrator harnesses a \textit{self-evolving reasoning mechanism} to propose candidate constructions, invoke purpose-built scientific tools for validation, and refine subsequent search strategies using the resulting evidence. In addition, to reconcile the large search space with the high cost of LER evaluation, \omni exploits a \textit{slow--fast synergistic workflow}. The fast loop generates and screens large populations of candidates using inexpensive code-level metrics, whereas the slow loop compiles selected candidates into noisy circuits, evaluates their LERs using configurable decoders, and feeds the resulting evidence back into both candidate generation and search refinement. In doing so, \omni incorporates improved physical models and evaluation methods without restructuring the discovery loop. Together, these components deliver an efficient, self-evolving AI scientist for physically grounded code--circuit--decoder co-design.

We systematically evaluate \omni under the practical constraints of deploying qLDPC codes on modern quantum processors. Specifically, we impose a fixed total physical-qubit budget $N$ and optimize the circuit-level LER over complete code implementations. Across four qLDPC construction families and three LLM backends (i.e., Claude, GPT, and DeepSeek), the scaling analysis shows that \omni discovers codes with progressively stronger error suppression as $N$ increases, implying an approximately linear resource--performance trend. The discovered codes outperform the standard bivariate-bicycle (BB) code with $[\![72,12,6]\!]$ when $N=98$~\cite{bravyi2024high}, while the strongest Claude Opus~4.8 campaign surpasses the BB benchmark with $[\![144,12,12]\!]$ at  $N=240$. Besides, the achieved results reveal that the widely used code-level figure of merit $\Phi_{\rm{code}}=kd^2/n$~\cite{cruz2026evolutionary,bravyi2024high,liang2025generalized,bravyi2010tradeoffs} does not always align with the circuit-level LER ranking. This discrepancy suggests a fundamental limitation of automated approaches that optimize code-level objectives alone.  Further evaluations show that the discovered codes retain their performance advantage across physical error rates beyond the single value used during search. These results establish \omni as a physically grounded framework for discovering qLDPC codes that admit resource-efficient processor-level implementations.

\section{Implementation of \omni}
Practical QEC discovery centers on bridging the gap between where candidates are generated and where their performance is ultimately determined~\cite{zhou2025opportunities}. This remains a long-standing challenge, because physical feedback is often delayed, expensive, and contingent on hardware-specific circuit realizations~\cite{baspin2022connectivity,strikis2026high,kishony2026surface,bi2026untangling,roffe2020decoding}, leaving conventional code-centric search poorly aligned with the final objective. Progress towards addressing this challenge could advance fault-tolerant quantum computing by accelerating the identification of practically competitive architectures.

\subsection{\omni system architecture}

\omni is designed to bridge this gap by integrating code discovery, syndrome-extraction circuit synthesis, and decoder-based evaluation within a unified workflow. Following the broader paradigm of AI-scientist systems~\cite{boiko2023autonomous,lu2024ai, chen2026agentic,m2024augmenting,arlt2026meta,kim2026capable}, it leverages the reasoning capabilities of LLMs and the tool-use capabilities of agentic systems to identify practical QEC codes under user-specified objectives~\cite{novikov2025alphaevolve}. At its core, an LLM-based orchestrator interprets the user’s objective, formulates a structured discovery task, and coordinates agents and computational tools throughout the discovery process. To support this discovery, we develop a suite of QEC-specific agents and tools for evaluating code properties, validating circuit-level performance, and exploring structured code families at scale.

An overview of \omni is shown in Fig.~\ref{fig:scheme}. The workflow begins by parsing a user-defined design objective. The LLM-based orchestrator interprets the query and translates it into a structured QEC discovery task, e.g., the target code family, optimization criteria, and relevant hardware constraints. These constraints may specify the total physical-qubit budget $N$, qubit connectivity, native gate set, available ancilla resources and other implementation requirements. After the interpretation, \omni initiates a self-evolving discovery process over $T$ iterations. At iteration $t\in[T]$, candidate code-construction programs are generated, evaluated, and refined by the orchestrator using feedback from previous iterations. The discovery process terminates when the target objective is met, the iteration limit $T$ is reached, or the available computational budget is exhausted.

\smallskip
\noindent{\underline{\textit{Dual-loop strategy.}}} A computational bottleneck in this iterative process is circuit-level evaluation, which requires constructing noisy syndrome-extraction circuits and estimating the corresponding logical performance through a large number of decoding simulations~\cite{gidney2021stim}. To address this bottleneck, we devise a \textit{dual-loop strategy} within each iteration, as shown in the middle and right panels of Fig.~\ref{fig:scheme}. \omni uses efficiently computed code-level properties as proxies to explore and refine a broad set of candidate constructions, while selectively allocating costly circuit-level evaluations to the most promising ones. The resulting circuit-level evidence, together with the associated reasoning trajectories~\cite{yao2022react,shinn2023reflexion,yan2026ai}, is then fed back to the orchestrator, enabling it to focus subsequent iterations on promising regions of the design space. In what follows, we present the implementation of this dual-loop discovery process and the agents and tools that support it. Further details are provided in Supplementary Information (SI)~A~and~B.

The dual-loop strategy at each iteration contains a fast loop and a slow loop. In the fast loop, \omni employs a code-search agent to propose executable construction programs for QEC codes with parameters $[\![n,k,d]\!]$ across the specified code families, such as lifted-product~\cite{panteleev2021asymptotically}, BB~\cite{bravyi2024high}, and multivariate-bicycle codes~\cite{voss2025multivariate}. Each program is assessed by the code-characterization tools, which verify the validity of the resulting code, and compute code-level properties such as the number of logical qubits $k$, code distance $d$, and the selected figure of merit $\Phi_{\rm code}=kd^2/n$~\cite{bravyi2024high,cruz2026evolutionary}. For Calderbank–Shor–Steane (CSS) codes~\cite{calderbank1996good,steane1996multiple}, specified by the binary $X$- and $Z$-check matrices $H_X$ and $H_Z$, the validity includes verifying the commutation condition $H_XH_Z^{\mathsf T}=0$ over $\mathbb{F}_2$~\cite{gottesman1997stabilizer}. These evaluations take only seconds per candidate, allowing a large set of constructions to be rapidly evaluated and ranked. The orchestrator selects the top-ranked candidates for the slow loop.

In the slow loop, each prioritized code from the fast loop is first passed to the circuit-synthesis module, which constructs and optimizes a syndrome-extraction circuit under the specified connectivity, native gate set, and scheduling constraints. The resulting circuit is then supplied to the circuit-evaluation module, which derives the corresponding detector error model~\cite{gidney2021stim}, generates noisy syndrome samples, and applies the selected decoders to estimate the circuit-level logical error rate (LER)~\cite{fowler2012surface}. This stage requires repeated noisy sampling and decoding~\cite{gidney2021stim,mayer2025rare}, which typically takes minutes to hours and is therefore reserved for a selected subset of candidates. The resulting circuit-level ranking, together with the associated synthesis, decoding, and reasoning records, is fed back to the orchestrator and incorporated into the discovery memory. The orchestrator uses this feedback to account for implementation-level performance and generate improved candidates in subsequent iterations.

\subsection{Construction of key components} 
The capabilities of \omni are determined by its constituent agents and tools. Here, we present the key components implemented in the current framework, which are tailored to QEC discovery for \textit{quantum-memory experiments}~\cite{google2023suppressing,google2025quantum,bravyi2024high}. Refer to SI~B for more details. As a flexible framework, its modular design allows additional agents and tools to be incorporated for other QEC tasks, such as the optimization of logical operations~\cite{zhang2025time,cain2024correlated}, lattice-surgery protocols~\cite{horsman2012surface, litinski2019game, chamberland2022universal}, and other fault-tolerant primitives~\cite{bravyi2005universal,bombin2015single,gidney2019efficient}.

\begin{figure*}[t!]
    \centering
    \includegraphics[width=\linewidth]{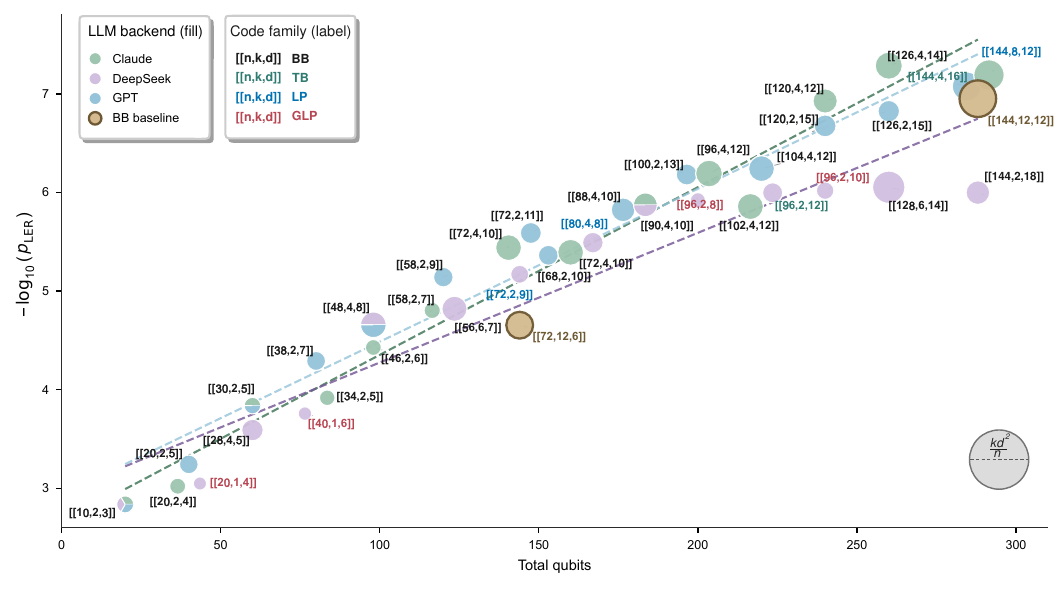}
 \caption{\textbf{Circuit-level performance frontier discovered by \omni.} Circuit-level performance as a function of the physical-qubit budget $N$, including data qubits and stabilizer-measurement ancillas. The vertical axis reports logical-error suppression, $-\log_{10}(p_{\rm LER})$, where $p_{\rm LER}$ is the logical error rate per round per logical qubit. Marker fill denotes the LLM backend used by the orchestrator, and pie markers indicate codes independently discovered by multiple backends. The BB $[\![72,12,6]\!]$ and $[\![144,12,12]\!]$ codes are included as reference codes. Label color denotes the code family, and circle area is proportional to the fast-loop code-level proxy $\Phi_{\rm code}$. The discovered codes form a clear resource--performance frontier, with several candidates achieving stronger circuit-level suppression than the BB reference codes despite having smaller $\Phi_{\rm code}$. This mismatch shows that code-level proxies can mis-rank operational performance and motivates circuit-level LER as the direct discovery objective.}
    \label{fig:mainresults}
\end{figure*}

\smallskip   
\noindent\textit{\underline{Orchestrator.}} The orchestrator serves as the control layer of the \omni framework. It translates a high-level design objective into a structured specification of the target code space, resource constraints and evaluation protocol, and routes tasks between the search agent and the scientific modules~\cite{boiko2023autonomous,romera2024mathematical,lu2024ai}. 

This functionality is realized using a general-purpose LLM backbone, such as GPT, Claude, or DeepSeek, equipped with task-specific prompts, tool-use interfaces, and access to the accumulated discovery history. For each candidate construction in each iteration, the orchestrator maintains a unified record containing its construction program, code-level properties, compiled circuits and decoding results. It distils this accumulated evidence into an evolving search context, enabling each iteration to generate better-informed constructions and progressively refine the discovery trajectory.

\smallskip
\noindent\textit{\underline{Code-search agent.}} This agent operates within the fast loop and proposes diverse executable QEC construction programs through LLM-guided program search~\cite{boiko2023autonomous,romera2024mathematical,lu2024ai}. We implement this agent using a dedicated LLM backbone, such as GPT or DeepSeek. It receives the structured design objective and evolving search context from the orchestrator, and draws on QEC-specific knowledge of code constructions, algebraic constraints, and established design principles to formulate and revise \textit{candidate programs}. Here each program refers to a construction rule together with its tunable parameters. For a CSS construction, for example, the rule may define how the binary check matrices $H_X$ and $H_Z$ are generated, while the tunable parameters control properties such as matrix size, polynomial support, sparsity pattern, and random seed~\cite{panteleev2021quantum,bravyi2024high,voss2025multivariate}. A single program can therefore generate a family of related codes across different parameter choices. This programmatic representation preserves the provenance of each candidate and enables promising constructions to be systematically refined across iterations.  

\smallskip
\noindent\textit{\underline{Code-characterization module.}} This module, as another key component in the fast loop, provides a general verification and evaluation interface between program generation and circuit-level analysis. It executes each construction program, checks whether the resulting code satisfies the defining algebraic conditions of the specified code family, and enforces the user-specified resource constraints. For CSS codes, for example, validity includes verifying the commutation condition $H_XH_Z^{\mathsf T}=0$. Exact linear-algebra routines~\cite{gottesman1997stabilizer,roffe2019quantum} are then used to compute certified quantities such as the block length $n$, the number of logical qubits $k$, check weights, and qubit degrees. Quantities that are generally more difficult to determine exactly, most notably the code distance $d$, are estimated or bounded using in-house or open-source decoder-assisted tools, including BP+LSD~\cite{hillmann2025localized} and OSD~\cite{roffe2020decoding}. The modular interface also allows additional characterization tools to be incorporated as the framework evolves.

 \smallskip
\noindent\textit{\underline{Circuit-synthesis module.}} This module operates within the slow loop and maps each selected code to an executable syndrome-extraction implementation required for circuit-level evaluation. Guided by the orchestrator, it allocates data qubits and ancillas, prepares the encoded state, and determines the protocol for repeated syndrome-extraction rounds. The protocol specifies how each stabilizer measurement is decomposed into native operations and how the resulting interactions are ordered and scheduled. For CSS codes, this entails separate measurements of the $X$- and $Z$-type checks specified by $H_X$ and $H_Z$.

The circuit-synthesis module in \omni is implemented by integrating two complementary libraries. Within each syndrome-extraction round, the required two-qubit interactions are scheduled through edge colouring of the associated Tanner graph~\cite{zhang2026optimal}, producing conflict-free parallel circuit layers. The resulting schedule is then compiled into a complete Stim circuit~\cite{gidney2021stim} comprising state-preparation and reset operations, native entangling gates, ancilla measurements, detector definitions and logical-observable annotations. The module also returns circuit-level metrics, such as syndrome-extraction depth and two-qubit-gate count, to the orchestrator as feedback for subsequent discovery iterations.

\smallskip 
\noindent\textit{\underline{Circuit-evaluation module.}} This module estimates the logical performance of each synthesized implementation under a configurable circuit-level noise model. The physical error rate can be specified by the user or chosen automatically at a common operating point slightly below the estimated pseudo-threshold, placing the evaluation in the error-suppression regime while retaining a sufficient number of logical failures for tractable direct sampling~\cite{mayer2025rare}.

The implementation of this module consists of circuit sampling and modular decoding. For each synthesized implementation, the noisy Stim circuit~\cite{gidney2021stim} is converted into a detector error model, and sampled to generate detection events and logical-observable outcomes. These samples are then passed to a decoder-agnostic interface that can accommodate a broad range of decoding algorithms, including matching-based, belief-propagation-based, and neural decoders. In the current framework, \omni integrates BP+OSD~\cite{roffe2020decoding,panteleev2021degenerate}, BP+LSD~\cite{hillmann2025localized} and Relay-BP~\cite{muller2025improved}. The resulting LER estimates and associated decoding records define the circuit-level ranking, which is returned to the orchestrator.

\section{Experimental results} 
We apply \omni to the discovery of practical qLDPC codes under explicit hardware budgets. We focus on qLDPC codes because their finite-rate encoding and sparse parity checks offer a promising route to reducing the physical-qubit overhead of fault-tolerant quantum computation~\cite{breuckmann2021quantum,panteleev2022asymptotically,bravyi2024high}. Practicality is enforced by restricting the total number of physical qubits $N$ to a scale comparable to the BB code with $[\![144, 12, 12]\!]$, which is relevant to early fault-tolerant quantum processors~\cite{bravyi2024high,katabarwa2024early}. The discovery of new qLDPC codes that match or surpass leading existing constructions at the circuit level, as measured by lower LER $p_{\rm LER}$, could broaden the range of fault-tolerant architectures realizable under realistic hardware constraints and accelerate progress towards practical fault-tolerant quantum computing~\cite{campbell2017roads,bravyi2024high}.

\subsection{Experiment setup and hyperparameter settings}

We conduct systematic experiments to evaluate the capability of \omni to discover practical qLDPC codes against the leading BB reference codes with $[\![72,12,6]\!]$ and $[\![144,12,12]\!] $, whose circuit implementations require total physical-qubit budgets of $N=144$ and $N=288$,  respectively. Specifically, the orchestrator of \omni is instantiated with three API-accessed LLM backends, which are Claude Opus~4.8, DeepSeek-V4-Pro, and GPT-5.5. The code-search agent explores four qLDPC construction families, i.e., BB, trivariate-bicycle (TB), lifted-product (LP), and generalized LP (GLP) codes, as detailed in SI~A. We consider $14$ total physical-qubit budgets spanning $N=20$ to $N=288$.

The hyperparameter settings employed in \omni are fixed across all resource budgets and LLM backends. For each $N$, \omni is executed for $T=60$ dual-loop iterations. At each iteration, the code-search agent generates $24$ construction programs, each instantiated with $20$ random seeds, yielding up to $480$ candidate code instances before validity checking and deduplication. The resulting code instances are assessed in the fast loop using the code-level objective $\Phi_{\rm code}$, after which a subset of the highest-ranking candidates is selected adaptively for slow-loop evaluation. In the slow loop, these candidates are compiled into syndrome-extraction memory circuits and evaluated using BP+OSD under circuit-level depolarizing noise with physical error rate $p=0.005$. Unless stated otherwise, we report the LER per syndrome-extraction round and per logical qubit, i.e., $p_{\rm LER}$. Refer to SI~C for a detailed definition for $p_{\rm LER}$ and other experimental details.
 
\begin{figure*}[tb!]
    \centering
    \includegraphics[width=\linewidth]{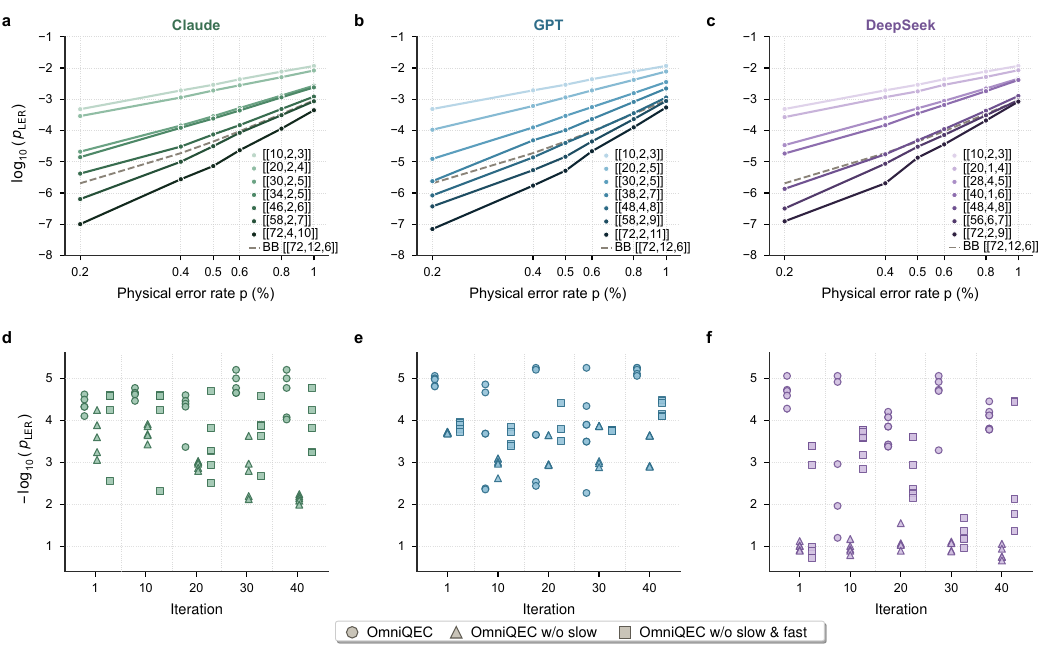}
\caption{\textbf{Circuit-level validation and loop ablation.}
\textbf{a--c,} Physical-error-rate sweeps for representative frontier codes discovered by \omni using Claude Opus~4.8 (\textbf{a}), GPT-5.5 (\textbf{b}) and DeepSeek-V4-Pro (\textbf{c}). The BB $[\![72,12,6]\!]$ code is shown as a reference under the same syndrome-extraction and decoding protocol. Solid lines denote \omni-discovered codes and dashed lines denote the BB reference. The fitted scaling $p_{\rm LER}\propto p^\Gamma$ gives the LER-scaling exponent $\Gamma$.
\textbf{d--f,} Ablation of discovery feedback for Claude Opus~4.8 (\textbf{d}), GPT-5.5 (\textbf{e}) and DeepSeek-V4-Pro (\textbf{f}). Each marker denotes one of the top-five candidates obtained at the corresponding iteration, evaluated by the same circuit-level protocol and plotted by its error suppression $-\log_{10}(p_{\rm LER})$. The full \omni workflow uses both fast-loop code-level screening and slow-loop circuit-level feedback. The fast-loop-only variant, denoted by \omni w/o slow, removes circuit-level feedback and ranks candidates using only code-level rankings. The no-feedback variant, denoted by \omni w/o slow \& fast, removes both feedback paths, so each iteration proceeds independently of previous results. The full workflow concentrates candidates in the high-suppression regime, whereas removing feedback increases variability and can steer the search towards circuit-level suboptimal candidates.}
    \label{fig:3}
\end{figure*}

\subsection{Scaling performance of \omni}
We first evaluate the performance of \omni across different physical-qubit budgets $N$, with the results illustrated in Fig.~\ref{fig:mainresults}. For all three LLM backends, the best discovered codes form a clear resource--performance frontier, with the LER $p_{\rm LER}$ decreasing approximately exponentially as $N$ increases. More specifically, all three backends surpass the BB $[\![72,12,6]\!]$ reference, with the first improvement occurring at $N=98$. This crossover point is particularly relevant because it lies within the hardware scale of current reconfigurable processors, such as Helios~\cite{ransford202698}. However, at larger physical-qubit budgets, the performance frontiers diverge across backends. DeepSeek yields the weakest frontier and does not surpass the BB $[\![144,12,12]\!]$ reference within the explored range, whereas both GPT and Claude exceed this reference at $N=240$. These differences highlight the role of the orchestrator backbone, as stronger reasoning and reflection capabilities enable circuit-level feedback to be converted more effectively into improved construction proposals.

A complementary analysis of the discovered code families, presented in SI~D, reveals distinct exploration behaviours across orchestrators. The leading candidates identified by Claude and GPT are dominated by BB constructions, whereas DeepSeek retains greater diversity across multiple code families but attains a weaker performance frontier. This contrast highlights an exploration–exploitation trade-off in \omni and suggests a direction for future improvement. That is, broad exploration preserves structural diversity, whereas effective exploitation depends on the orchestrator recognizing when circuit-level evidence is sufficiently consistent to focus subsequent search on high-performing code structures.

Another finding from the scaling analysis is that the circuit-level performance frontier is poorly aligned with the fast-loop objective. The proxy $\Phi_{\rm code}$, which is widely adopted in previous code-centric code searches~\cite{cruz2026evolutionary,he2026co,liu2026large}, assigns scores of $6$ and $12$ to the BB $[\![72,12,6]\!]$ and $[\![144,12,12]\!]$ references, respectively. Yet only one code discovered by \omni exceeds the larger reference under this metric. Despite their lower proxy scores, many codes discovered at $N\geq 98$ achieve lower circuit-level LER than the BB $[\![72,12,6]\!]$ reference. The mismatch is especially evident at $N=144$, where the best codes found with DeepSeek, Claude and GPT have proxy scores of $\Phi_{\rm code}=2.25$, $5.56$ and $3.36$, respectively, while all outperform the BB reference in circuit-level LER. This rank mismatch underlines the need to use LER as a direct objective in practical QEC code design, beyond relying on code-level proxies alone.

\begin{figure}[t]
    \centering
    \includegraphics[width=0.6\linewidth]{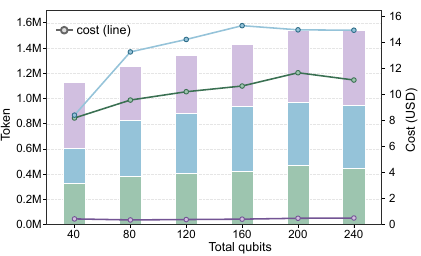}
   \caption{\textbf{LLM-side token usage and API cost.}
Bars show total API token consumption, including input, output and reasoning tokens, and lines show the corresponding API cost for each backend. Colors denote the LLM backend, using the same backend color coding as in Fig.~\ref{fig:3}. }
    \label{fig:4}
\end{figure}

\subsection{Circuit-level validation of \omni}
We next test whether the advantage of the discovered codes persists across physical error rates $p$. To this end, we select seven frontier codes discovered by \omni for each LLM backend and perform full physical-error-rate sweeps, yielding 21 codes in total. We re-evaluate these codes and the BB $[\![72,12,6]\!]$ reference over $p\in[0.002,0.01]$ using identical syndrome-extraction and decoding protocols. As shown in Fig.~\ref{fig:3}a--c, the discovered codes maintain lower LER than the BB reference throughout the measured range. At the two endpoints, the strongest candidate achieves an LER that is $29.23$-fold lower at $p=0.002$ and $1.66$-fold lower at $p=0.01$.

We further quantify the dependence of the LER $p_{\rm LER}$ on $p$ by fitting $p_{\rm LER}\propto p^{\Gamma}$, where $\Gamma$ is the fitted LER-scaling exponent. The representative codes, i.e., $[\![48,4,8]\!]$ discovered with GPT, $[\![58,2,7]\!]$ discovered with Claude and $[\![48,4,8]\!]$ discovered with DeepSeek, yield $\Gamma=5.20$, $5.60$ and $5.57$, respectively. Compared with $\Gamma=3.79$ for the BB reference, these results show that the advantage of the \omni-discovered codes persists beyond the single physical error rate used during discovery.

\subsection{Ablation study of \omni}
We then move on to comprehend the respective roles of the fast and slow loops through ablation experiments. To be concrete, we compare the full \omni workflow with two reduced variants. The fast-loop-only variant, denoted by \omni w/o slow, removes the circuit-level feedback, so the discovery process is solely based on code-level rankings. The no-feedback variant, denoted by \omni w/o slow \& fast, removes both feedback paths, causing each iteration to proceed independently of previous results. For a fair comparison, the top five candidates from each variant are evaluated using the same circuit-level protocol. 

The achieved results are demonstrated in Fig.~\ref{fig:3}d--f. The full $\omni$ workflow consistently attains stronger and more stable error suppression than its ablated variants. At the final iteration, the top-five candidates reach mean error suppressions of $-\log_{10}(p_\mathrm{LER}) = 4.61 \pm 0.54$, $5.17 \pm 0.09$, and $4.07 \pm 0.28$ for Claude Opus~4.8, GPT-5.5, and DeepSeek-V4-Pro, respectively. Compared with the full workflow, $\omni$ w/o slow reaches only $2.14 \pm 0.10$, $3.20 \pm 0.40$, and $0.84 \pm 0.16$, while $\omni$ w/o slow \& fast reaches $3.86 \pm 0.66$, $4.25 \pm 0.19$, and $2.83 \pm 1.50$. Across the three backends and all checkpoints, the full workflow further achieves the highest mean suppression, $4.27$, and the lowest variance, $0.79$. Removing feedback increases the variance substantially to $1.24$--$1.26$. Taken together, these results show that the fast loop improves search efficiency, whereas the slow loop is essential for steering discovery towards practically relevant circuit-level performance.
 
\subsection{Implementation cost of \omni}
We lastly assess the LLM-side cost of \omni across different backends. As shown in Fig.~\ref{fig:4}, API token usage varies only weakly with the physical-qubit budget $N$, indicating that larger code instances do not substantially increase the required context. When considered alongside the achieved performance frontiers, Claude offers the most favourable cost–performance balance, whereas DeepSeek is the least expensive but produces the weakest frontier. These results indicate that LLM-side cost is governed primarily by backend-specific reasoning behaviour and API pricing, rather than by the size of the codes being explored.

\section{DISCUSSION}
In this study, we introduce \omni, a new paradigm for automated QEC code discovery. Unlike prior approaches that optimize code-level proxies, \omni directly targets implementation-level performance. To make this objective computationally tractable, \omni harnesses a purpose-built self-evolving reasoning mechanism, specialized scientific tools and a slow–fast synergistic workflow. \omni thereby provides a scalable route towards discovering practical, high-performing QEC codes.

Our study points to several broader directions for autonomous QEC design. A first direction is to develop more capable and efficient scientific tools for \omni. Its main computational bottleneck is circuit-level verification, which limits the number of candidates that can be evaluated using physical evidence. One potential route is to improve key components such as syndrome-extraction synthesis~\cite{zhang2026optimal,viszlai2026prophunt}, circuit simulation and decoding~\cite{gidney2021stim,senior2025scalable,yan2026rethink,yan2026efficient}, together with more efficient statistical estimation of logical performance. Because these components are exposed through modular interfaces, new tools could be integrated directly into \omni, enabling broader searches over richer and more hardware-relevant objectives~\cite{senior2025scalable,yan2026rethink,yan2026efficient}.

Moreover, it is important to extend \omni from its present agent--module architecture towards an end-to-end multi-agent system~\cite{li2023camel,zhang2025multi,zhang2025landscape}. At present, only the orchestrator and code-search agent use LLM backends, whereas code characterization, circuit synthesis, and circuit evaluation invoke predefined tools. As the toolset grows, these modules could be upgraded into specialized agents that adapt circuit schedules, flag-qubit placement and decoder configurations to each candidate code. This would enable increasingly autonomous co-design of codes, circuits and decoders, while keeping evaluation grounded in executable tools and verifiable evidence~\cite{zheng2023judging,dsouza2025yescieval,findeis2025external}.

Another important direction is to extend \omni towards broader hardware objectives and code families.  Recent work has shown that QEC detection events can provide learning signals for continuously adjusting hardware-control parameters~\cite{sivak2026reinforcement}. In \omni, calibration data such as gate-error maps and disabled qubits or couplers could instead guide code selection, physical embedding, syndrome-extraction scheduling, and decoder configuration. This would extend the present hardware-aware search towards calibration-aware code--circuit--decoder co-design for a specific processor. The search space could also expand beyond the four families studied here to higher-dimensional hypergraph-product codes~\cite{zeng2019higher}, two-block group-algebra codes~\cite{lin2024twoblock}, balanced-product codes~\cite{breuckmann2021balanced} and Floquet codes~\cite{hastings2021dynamically,davydova2023floquet}. Together, richer objectives and construction templates would enable \omni to discover codes tailored to neutral-atom arrays, trapped-ion processors, and other reconfigurable platforms~\cite{bluvstein2024logical,xu2024constant,egan2021fault,ryananderson2021realization}.

A parallel direction is to advance the entire \omni pipeline towards a hierarchical multi-agent system, in which specialized agents coordinate under a central scientific orchestrator~\cite{li2023camel,zhang2025multi,zhang2025landscape}. For example, a circuit-synthesis agent could compare candidate schedules and optimize circuit depth or ancilla overhead, whereas a decoding agent could select decoder families and tune their configurations for each code. An independent evaluation agent could further assess intermediate proposals, although the known biases of LLM-as-a-judge approaches motivate grounding such assessments in executable tools and verifiable physical evidence~\cite{zheng2023judging,dsouza2025yescieval,findeis2025external}. Such an architecture could extend \omni from tool-assisted code discovery to increasingly autonomous code–circuit–decoder co-design.

\end{document}